\documentclass[aps,prx,twocolumn,footinbib,superscriptaddress,showpacs,preprintnumbers]{revtex4-2}

\usepackage[utf8]{inputenc}

\usepackage{cancel}
\usepackage{comment}

\usepackage[dvipsnames]{xcolor}      
\definecolor{lcolor}{rgb}{0.5,0,0}
\definecolor{citcolor}{rgb}{0,0.0,1}

\usepackage[breaklinks,colorlinks,urlcolor=blue,citecolor=citcolor,linkcolor=lcolor,linktoc=all]{hyperref}
\usepackage{color}
\usepackage{graphicx}	
\graphicspath{{./figures/}}
\usepackage[utf8]{inputenc}
\usepackage{amsmath}
\usepackage{amssymb}
\usepackage{amsfonts}
\usepackage[ragged]{footmisc}
\usepackage{slashed}
\usepackage{float}
\usepackage[capitalise]{cleveref}
\usepackage{dsfont}
\usepackage{enumerate}

\usepackage[normalem]{ulem}

\makeatletter
\g@addto@macro\bfseries{\boldmath}
\makeatother

    \DeclareFontFamily{U}{wncy}{}
    \DeclareFontShape{U}{wncy}{m}{n}{<->wncyr10}{}
    \DeclareSymbolFont{mcy}{U}{wncy}{m}{n}
    \DeclareMathSymbol{\Sh}{\mathord}{mcy}{"58}

\usepackage{tikz}
\usepackage[customcolors]{hf-tikz}

\usetikzlibrary{arrows,cd,shapes,decorations.pathmorphing,decorations.markings,shadings}
\tikzset{
  big arrow/.style={
    decoration={markings,mark=at position 1 with {\arrow[scale=4,#1]{>}}},
    postaction={decorate},
    shorten >=0.4pt},
  big arrow/.default=blue}

\newcommand{\bbone}{\text{\usefont{U}{bbold}{m}{n}1}}
\MakeRobust{\bbone}

\newcommand{\Lbar}{\overline{\Lambda}}
\newcommand{\mE}{m_{\text{E}}}

\renewcommand{\epsilon}{\varepsilon}

\usepackage[acronym]{glossaries}
\newacronym{LO}{LO}{leading-order}

\newacronym{QM}{QM}{quark matter}

\newacronym{QCD}{QCD}{quantum chromodynamics}

\newacronym{HTL}{HTL}{Hard-Thermal-Loop}

\newacronym{UV}{UV}{ultraviolet}

\newacronym{IR}{IR}{infrared}

\begin{document}

\title{Complete four-loop pressure of cold QED}

\preprint{CERN-TH-2026-251, HIP-2026-18/TH}

\author{Aapeli Kärkkäinen}
\email{aapeli.karkkainen@helsinki.fi}
\affiliation{Department of Physics and Helsinki Institute of Physics, P.O.~Box 64, FI-00014 University of Helsinki, Finland}
\author{Pablo Navarrete}
\email{pablo.navarrete@helsinki.fi}
\affiliation{Department of Physics and Helsinki Institute of Physics, P.O.~Box 64, FI-00014 University of Helsinki, Finland}
\author{Mika Nurmela}
\email{mika.nurmela@helsinki.fi}
\affiliation{Department of Physics and Helsinki Institute of Physics, P.O.~Box 64, FI-00014 University of Helsinki, Finland}
\author{Kaapo Seppänen}
\email{kaapo.seppaenen@unibe.ch}
\affiliation{AEC, Institute for Theoretical Physics, University of Bern, Sidlerstrasse 5, CH-3012 Bern, Switzerland}
\author{\\Samuel Abreu}
\email{samuel.abreu@ugr.es}
\affiliation{Theoretical Physics Department, CERN, CH-1211 Geneva 23, Switzerland}
\affiliation{Higgs Centre for Theoretical Physics, School of Physics and Astronomy, The University of Edinburgh, Edinburgh EH9 3FD, Scotland, United Kingdom}
\affiliation{Departamento de Fisica Teorica y del Cosmos, Universidad de Granada, Campus de Fuentenueva, E-18071 Granada, Spain}
\author{Maximilian Delto}
\email{maximilian.delto@cern.ch}
\affiliation{Theoretical Physics Department, CERN, CH-1211 Geneva 23, Switzerland}
\author{Aleksi Kurkela}
\email{aleksi.kurkela@uis.no}
\affiliation{Faculty of Science and Technology, University of Stavanger, NO-4036 Stavanger, Norway}
\author{Risto Paatelainen}
\email{risto.paatelainen@helsinki.fi}
\affiliation{Department of Physics and Astronomy, FI-20014 University of Turku, Finland}
\affiliation{Department of Physics and Helsinki Institute of Physics, P.O.~Box 64, FI-00014 University of Helsinki, Finland}
\author{Aleksi Vuorinen}
\email{aleksi.vuorinen@helsinki.fi}
\affiliation{Department of Physics and Helsinki Institute of Physics, P.O.~Box 64, FI-00014 University of Helsinki, Finland}

\begin{abstract}

Extending the weak-coupling expansion of the pressure of a cold and dense ultrarelativistic electron gas to next-to-next-to-next-to-leading order (N3LO) in the gauge coupling has remained an open problem since the classic N2LO result of Freedman and McLerran, published in 1977.
Here, we present the full N3LO pressure, the first thermodynamic quantity determined to this order in a physical gauge theory.
The pressure exhibits good perturbative convergence and weak renormalization-scale dependence even at sizable couplings, while a Bayesian analysis of missing higher-order contributions indicates uncertainties that can exceed those inferred from scale variation alone.
\end{abstract}

\maketitle

\section{Introduction}

The past two decades have witnessed remarkable progress in perturbative quantum field theory calculations in vacuum, driven by advances in multiloop techniques. 
In Quantum Chromodynamics (QCD), for instance, many fundamental quantities are now known to five-loop precision, e.g.~\cite{Baikov:2016tgj,Herzog:2017ohr,Luthe:2016xec,Luthe:2017ttc}. 
A crucial tool facilitating this progress has been the use of integration-by-parts (IBP) reductions~\cite{Tkachov:1981wb,Chetyrkin:1981qh,Laporta:2000dsw}, which reduce the number of loop integrals one needs to evaluate. 
These calculations are, however, famously plagued by singularities in intermediate expressions, which require the use of dimensional regularization. 
To make cancellations among these singularities manifest early on, new ways of organizing perturbative calculations have been developed, including the loop-tree duality (LTD)  approach \cite{Catani:2008xa,Bierenbaum:2010cy,Capatti:2019ypt,Runkel:2019yrs,Capatti:2020ytd,JesusAguilera-Verdugo:2020fsn,Aguilera-Verdugo:2020set,Capatti:2022mly}.

In comparison, the development of perturbative thermal field theory --- including both nonzero temperatures and densities --- has progressed at a considerably slower pace.
Although so-called hard contributions to thermodynamic observables, entering through loop expansions, remain closely related to ordinary vacuum amplitudes especially for cold and dense systems \cite{Ghisoiu:2016swa}, the presence of a medium alters both the analytic structure of propagators and the organization of perturbation theory.
With loop integrals becoming intertwined with thermal sums or finite-density phase-space constraints, and the equilibrium state no longer invariant under Lorentz boosts, many techniques developed for vacuum amplitudes, such as IBP reduction, no longer apply straightforwardly.
At the same time, the screening of soft gauge-field modes by the medium generates parametrically separated momentum scales, causing the naive loop expansion to break down at low orders.
Their systematic treatment requires resummations or, equivalently, the use of an appropriate soft effective field theory (EFT) framework.
For a long time, these complications obstructed the extension of the perturbative pressure of cold and dense relativistic matter beyond the classic next-to-next-to-leading-order (N2LO) result of Freedman and McLerran \cite{Freedman:1976dm,Freedman:1976ub}.

Over the past decade, substantial progress has been achieved in overcoming these obstacles.
In particular, systematic work on the Hard Thermal Loop (HTL) effective theory has led to a detailed understanding of the infrared (IR) structure of cold and dense matter both for QCD and Quantum Electrodynamics (QED).
This has enabled a separation of the hard-, soft-, and mixed-scale contributions to thermodynamic observables, leading to a complete determination of the soft and mixed terms in the N3LO pressures of QED and QCD \cite{Kurkela:2016was,Gorda:2022zyc,Gorda:2018gpy,Gorda:2021znl,Gorda:2023mkk,Raza:2026imv}.
The only ingredient missing from the corresponding equations of state (EoS) is the hard contribution, arising from four-loop vacuum diagrams evaluated in the presence of a nonzero chemical potential $\mu$.
Very recently, this sector has also witnessed progress, with the vacuum LTD method generalized to nonzero chemical potentials~\cite{Navarrete:2024zgz} and applied to a subset of the necessary four-loop diagrams~\cite{Karkkainen:2025nkz} (see also the $T\neq 0$ extensions \cite{Navarrete:2025yxy,Seppanen:2025owq,Gorda:2025cwu}).

In this work, we complete this program for cold and dense massless QED, formulated in a jellium setup with a non-dynamical positive background charge density rendering the medium electrically neutral \cite{GiulianiVignale2005} --- a reasonable approximation to the electron-ion matter found inside massive white dwarfs.
Our result constitutes the first N3LO determination of a thermodynamic quantity in a physical gauge theory \footnote{Note, however, the four-loop computation of the pressure of high-temperature $\phi^4$ theory \cite{Gynther:2007bw}.} and provides a significant subset of the integrals required to compute the pressure of cold and dense quark matter at the same order.

\begin{figure*}[t!]
    \includegraphics[width=0.8\textwidth]{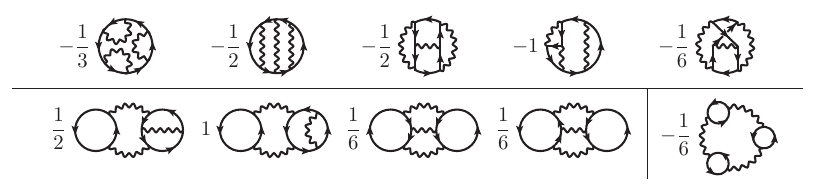}
\caption{All four-loop 1PI vacuum diagrams of QED, shown with their signs and symmetry factors.}
\label{fig:diags}
\end{figure*}

\section{Setup and Computation}

Through $O(\alpha_e^3)$, the weak-coupling expansion of the QED pressure receives contributions from three distinct momentum regions: The hard regime, where all loop momenta are of order $\mu$; the soft one, with photon momenta of order the electric screening mass $\mE\sim\sqrt{\alpha_e}\mu$; and the so-called mixed sector that involves both scales \cite{Gorda:2021znl,Gorda:2021kme}.
Accordingly, the pressure can be decomposed as
\begin{eqnarray}
\label{eq:pQED}
p&=& p^\text{h} + p^\text{s} + p^\text{m},
\end{eqnarray}
where 'h', 's', and 'm' stand for 'hard', 'soft', and 'mixed', and the corresponding contributions start at $O(\alpha_e^0)$, $O(\alpha_e^2)$, and $O(\alpha_e^3)$, respectively. While the hard and mixed terms are expected to contribute at all higher orders in $\alpha_e$, soft (HTL-resummed) photons do not self-interact \cite{Bellac:2011kqa}, implying that $p_\text{QED}^\text{s}$ is one-loop exact within the HTL effective theory and hence proportional to $\alpha_e^2$.

The mixed contribution at $O(\alpha_e^3)$ was first determined in \cite{Gorda:2022zyc} and later extended to QCD in \cite{Gorda:2023mkk} using dimensional regularization in $D \equiv d + 1 = 4-2\varepsilon$ dimensions.
The renormalized result takes the form
\begin{equation}
\label{eq:pmixed}
p^\text{m} = \frac{\alpha_e\mE^4}{(4\pi)^3} \left (\frac{\mE}{\Lbar} \right )^{-2\varepsilon}\left  [\frac{p^m_{-1}}{2\varepsilon} + p^m_{0} + O(\varepsilon) \right ],
\end{equation}
where $\alpha_e \equiv \alpha_e(\Lbar)$ is the renormalized coupling in the $\overline{\text{MS}}$ scheme and we have identified the factorization scale separating the hard and soft momentum regions with the renormalization scale $\Lbar$ \cite{Gorda:2021znl}.

Specializing to $N_f$ fermion flavors with common chemical potential $\mu$, the leading-order screening mass is $\mE^2 = 4\mu^2N_f\alpha_e(\Lbar)/\pi$ in $d=3$ (we use its $d$-dimensional form throughout \cite{Gorda:2022fci}), while the coefficients in eq.~\eqref{eq:pmixed} read
\begin{equation}
\begin{split}
    p^m_{-1} &= -11+\frac{5\pi^2}{6}+\left( \frac{5}{6}-\frac{7\pi^2}{72}-\frac{4}{3}\,\ln X \right)N_f, \\
    p^m_{0} &= -3.7988-2.7753\,\ln X \\  & \quad +\biggl (0.13787
    -1.3555\,\ln X-\frac{2}{3}\,\ln^2 X\biggr )N_f .
\end{split}
\label{eq:mixedcoefficients}
\end{equation}
Here, $X \equiv \Lbar/(2\mu)$, while the numerical constants shown admit one-dimensional integral representations that can be easily evaluated to arbitrary precision.
These results differ from those reported in \cite{Gorda:2022zyc} due to an $O(\varepsilon)$ normalization error in that reference, which propagates into $p^\text{m}_0$.
We have verified that eqs.~\eqref{eq:pmixed} and \eqref{eq:mixedcoefficients} fully agree with the Abelian limit of the QCD result \cite{Gorda:2023mkk}, providing an independent check of the corrected expression.

Turning to the hard contribution at N3LO, the corresponding term in eq.~\eqref{eq:pQED} is given by the sum of the 10 one-particle-irreducible (1PI) four-loop vacuum diagrams of QED. These graphs are shown in fig.~\ref{fig:diags}, grouped according to their number of fermion loops, each of which contributes one power of $N_f$ to the result.
The unique $O(N_f^3)$ diagram was already evaluated in \cite{Gorda:2022zyc} by exploiting its simple structure in terms of the one-loop photon self-energy.
The last two $O(N_f^2)$ diagrams in fig.~\ref{fig:diags} are on the other hand known to be sensitive to the complex phase of the fermion determinant \cite{Moore:2023glb}, with the corresponding contribution reducing to a pure number in the Abelian case \cite{Navarrete:2024zgz} (see appendix~\ref{app:example}).
These partial results provide non-trivial cross-checks for our computational method, which we apply to all graphs contributing to the pressure.

Our evaluation of the vacuum diagrams follows the strategy developed for QCD in \cite{Karkkainen:2025nkz} (see also \cite{Navarrete:2024ruu}).
We first use \texttt{Qgraf} \cite{Nogueira:1991ex} to generate the diagrams, then apply Feynman rules in a general covariant $R_\xi$ gauge, and subsequently perform the resulting contractions and algebraic manipulations using \texttt{FORM} \cite{Davies:2026cci}.
This yields a linearly dependent set of $O(10^5)$ scalar four-loop integrals with up to nine propagators and a varying number of scalar products of momenta in the numerator.
Using linear transformations of loop momenta with unit Jacobian, we algorithmically reduce each integrand to a canonical form \cite{Navarrete:2024ruu}, eliminating most redundancies and reducing the pressure to a sum of only 75 four-loop integrals (compared with 114 in QCD \cite{Karkkainen:2025nkz}), each multiplied by a polynomial coefficient in $d$. In what follows, we will refer to these integrals as master integrals~\footnote{We note that this terminology deviates from that used in the context of vacuum quantum field theory, where master integrals are irreducible under integration-by-parts relations; see e.g.~\cite{Smirnov:2010hn}}.
Adding all contributions, we find that the gauge dependence cancels, providing a consistency check of our calculation.

Further linear relations among our integrals can be derived using IBP identities, both in vacuum \cite{Chetyrkin:1981qh,Laporta:2000dsw} and in a medium \cite{Nishimura:2012ee,Osterman:2023tnt}.
For example, two-loop vacuum self-energy integrals can be simplified with well-known recursion relations \cite{Tarasov:1997kx}, collapsing certain four-loop integrals to trivial two-loop problems \cite{Osterman:2023tnt}.
We employ such relations as additional cross-checks of our results.

With the pressure reduced to a manageable set of integrals, we evaluate those accessible to semi-analytic methods using cutting-rule techniques \cite{Ghisoiu:2016swa,Kurkela:2009gj} and self-energy manipulations \cite{Arnold:1994ps,*Arnold:1994eb,Vuorinen:2003fs}.
For a subset of integrals, we also employ an independent approach~\cite{newapp} based on IBP reduction and differential equations \cite{Kotikov:1990kg,Remiddi:1997ny,Gehrmann:1999as}, which provides valuable cross-checks.
The remaining, more challenging integrals are evaluated numerically within the thermal LTD framework \cite{Navarrete:2024zgz,Karkkainen:2025nkz} (see also \cite{Seppanen:2025owq} for a detailed exposition), using local subtractions to handle ultraviolet (UV) and, where needed, IR divergences.
Together, these methods determine the required Laurent expansions up to $O(\varepsilon^0)$ in dimensional regularization.
Further details on the evaluation methods and their application to different integral topologies are given in appendix~\ref{app:masters} and in a forthcoming paper \cite{companion}.

Having expressed the bare hard pressure as a Laurent series in $\varepsilon$, we implement renormalization by expressing the bare coupling in terms of the renormalized one, including the counterterm contributions generated by lower-order terms.
Denoting the corresponding \text{renormalized} hard contribution by $p^\text{h,3}$, we obtain
\begin{equation}
\begin{split}
\label{eq:phard}
    p^\text{h,3} = \frac{\alpha_e \mE^4}{(4\pi)^3}\left(\frac{\Lbar}{\mu}\right)^{2\varepsilon} \left[ \frac{p^\text{h,3}_{-1}}{2\varepsilon}+p^\text{h,3}_0 +O(\varepsilon) \right],
\end{split}
\end{equation}
with the coefficients given by
\begin{equation}
\begin{split}
\label{eq:phardcoeffs}
    p^\text{h,3}_{-1} &= 11-\frac{5\pi^2}{6}+\left( -\frac{5}{6}+\frac{7\pi^2}{72}+\frac{4}{3}\,\ln X \right)N_f, \\
    p^\text{h,3}_{0} &= -2.833(24)\,N_f^{-1}-0.007(18)+1.60866\,\ln X \\
    &+ \left( 0.115547+0.561506\,\ln X+\frac{4}{9}\,\ln^2 X \right)N_f.
\end{split}
\end{equation}
These expressions constitute the main result of our paper.

Concentrating now on the renormalized pressure in eq.~\eqref{eq:phard}, we identify the remaining divergence as an IR one associated with soft photons in the QED medium. 
As can be readily verified, it exactly cancels the corresponding divergence in the mixed contribution of eq.~\eqref{eq:pmixed}, ensuring that the final result is finite.

While the pole term can be determined analytically, the evaluation of $p^\text{h,3}_0$ in eq.~\eqref{eq:phardcoeffs} requires computationally demanding Monte Carlo phase-space integrations within the LTD framework for roughly 70\% of our set of integrals, amounting to an estimated $10^6$ CPU-core hours (see appendix~\ref{app:masters} for details).
Uncertainties given in parentheses denote one-standard-deviation numerical integration errors, propagated to the pressure coefficients with correlations from shared master integrals retained.
For the $O(N_f^3)$ contribution and the sum of the last two $O(N_f^2)$ diagrams in fig.~\ref{fig:diags}, we use the more precise literature values \cite{Gorda:2022zyc,Navarrete:2024zgz}, which agree with our independent results.

 \par
 \nobreak
 \vspace*{0.5cm}
\section{The N3LO pressure: analysis of results}

With the mixed and hard pieces contributing to the pressure at $O(\alpha_e^3)$ now determined, it is a simple exercise to expand the resulting expressions in $\varepsilon$ and combine the terms to construct the physical pressure $p$.
Aside from the cancellation of the IR singularities in the sum $p^\text{h}+p^\text{m}$ that we already discussed, renormalization-group invariance through N3LO can also be readily verified as a further consistency check.
While performing this check, we identified a factor-of-two error in the handling of the N2LO result in \cite{Gorda:2022zyc}, related to the $O(N_f^{-1})$ contribution to $p_{2,3}$ (see table~\ref{tab:coeffs}), which led to an incorrect inferred value for $p_{3,4}$.

\begin{figure*}[th]
    \begin{center}
    \includegraphics[width=0.49\linewidth]{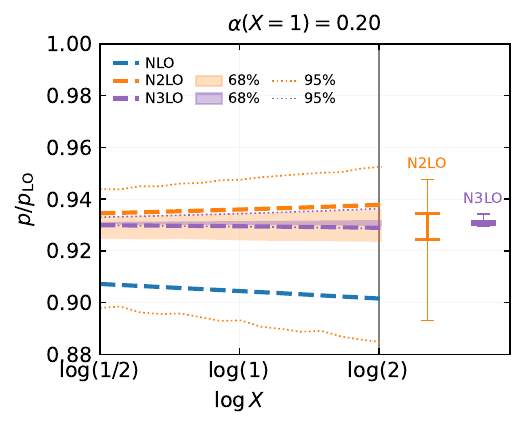}
    \includegraphics[width=0.49\linewidth]{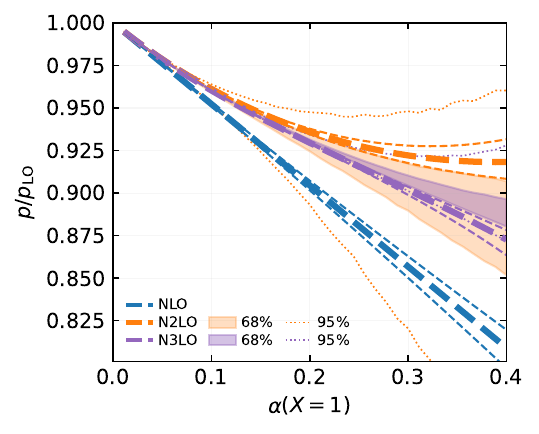}
    \caption{
    Left: The NLO (blue dashed line), N2LO (orange dashed line), and N3LO (purple dashed line) pressures of cold and dense QED, normalized by the free limit, as functions of the renormalization-scale parameter $X$ at the fixed coupling $\alpha_e(X=1)=0.2$.
    The 68\% and 95\% posterior credible intervals estimate the missing-higher-order uncertainty using the \emph{abc} model of Ref.~\cite{Duhr:2021mfd} and are conditioned separately at each $X$ using the result at the corresponding perturbative order.
    The isolated N2LO and N3LO bars show the final 68\% and 95\% uncertainty estimates, including both missing-higher-order and renormalization-scale uncertainties, with different values of $X$ combined using the scale-marginalization prescription of Ref.~\cite{Duhr:2021mfd}. \\
    Right: The pressure at different perturbative orders as a function of the coupling $\alpha_e(X=1)$.
    The thin dashed lines display the conventional renormalization-scale variation over $X\in[1/2,2]$, while the 68\% and 95\% posterior credible intervals account jointly for missing-higher-order and renormalization-scale uncertainties using the scale-marginalized \emph{abc} model.
    }
    \label{fig:results}
    \end{center}
\end{figure*}

Converting all coefficients to numerical form and normalizing by the leading-order result $p_\text{LO}=\mu^4 N_f/(12\pi^2)$, the \textit{complete} N3LO pressure of massless QED becomes
\begin{widetext}
\begin{equation}
\begin{split}
\label{eq:pfull}
    \frac{p}{p_\text{LO}}&=1+p_{1,1}\left( \frac{\alpha_e}{\pi} \right)+N_f\left( \frac{\alpha_e}{\pi} \right)^2\left[ p_{2,1}\ln\left( N_f\frac{\alpha_e}{\pi} \right)+p_{2,2}\ln X+p_{2,3} \right] \\
    &+ N_f^2 \left( \frac{\alpha_e}{\pi} \right)^3 \left[ p_{3,1}\ln\left( N_f\frac{\alpha_e}{\pi} \right) + p_{3,2} \ln\left( N_f\frac{\alpha_e}{\pi} \right) \ln X+p_{3,3} \ln^2X+p_{3,4} \ln X+p_{3,5} \right] + O(\alpha_e^4),
\end{split}
\end{equation}
\end{widetext}
with all coefficients duly listed in table~\ref{tab:coeffs}. 
The numerical uncertainties of our calculation are reflected in the $p_{3,5}$ coefficient reported in the table.

\begin{table}[!t]
\begin{tabular}{@{\quad}l@{\quad\quad\quad}l}
\toprule
    $p_{1,1}$  &  $-3/2$ \\
    $p_{2,1}$  &  $-3/2$ \\
    $p_{2,2}$  &  $1$ \\
    $p_{2,3}$  &  $0.45717+51/16\,N_f^{-1}$ \\
    $p_{3,1}$  &  $0.18932+4.16299\,N_f^{-1}$ \\
    $p_{3,2}$  &  $2$ \\
    $p_{3,3}$  &  $-2/3$ \\
    $p_{3,4}$  &  $0.39042-7/2\, N_f^{-1}$ \\
    $p_{3,5}$  &  $1.02270(2)-5.64(5)\,N_f^{-1}-8.50(7)\,N_f^{-2}$  \\
\botrule
\end{tabular}
\caption{Numerical values of the pressure coefficients appearing in eq.~\eqref{eq:pfull}. Decimal entries without an error estimate can be evaluated to arbitrary precision.}
    \label{tab:coeffs}
\end{table}

In the two panels of fig.~\ref{fig:results}, we display the behavior of the result for $N_f=1$ and explore its dependence on $\alpha_e$ rather than the chemical potential due to the small physical value of the fine-structure constant.
In order to assess the accuracy of our result, we first note that the numerical uncertainties in the pressure coefficients are negligible on the scale of fig.~\ref{fig:results} so they will be neglected in the analysis. 
We are thus left with two sources of uncertainties: The residual dependence on the unphysical renormalization scale $\bar{\Lambda}$ and the uncertainty associated with missing higher-order terms owing to truncation of the perturbative series.

Beginning from scale dependence, following historical conventions (see, e.g., \cite{Kurkela:2009gj}) we choose the central renormalization scale as $\bar{\Lambda}=2\mu$ and assess the associated scale dependence by varying $X\equiv\bar{\Lambda}/(2\mu)$ by a factor of two around unity, evolving $\alpha_e$ using the one-, two-, and three-loop QED beta functions at NLO, N2LO, and N3LO, respectively.
The left panel of fig.~\ref{fig:results} displays these three results at a fixed (large) coupling $\alpha_e(X=1)=0.2$, revealing that the differences between successive orders are significantly larger than the scale dependence.
Thus, conventional factor-of-two scale variation alone does not capture the observed order-to-order shifts.

To estimate the missing-higher-order uncertainty, we follow the Bayesian approach introduced in Ref.~\cite{Cacciari:2011ze}.
Specifically, we model the perturbative series using the asymmetric \emph{abc} model of Ref.~\cite{Duhr:2021mfd} (see also Ref.~\cite{Gorda:2023usm} for an application to finite-density QCD).
In this framework, the perturbative coefficients are treated as independent draws from a common statistical distribution, the parameters of which are inferred from the known coefficients, and the resulting posterior distribution is used to estimate the first unknown term in the series as well as the associated missing-higher-order uncertainty.

The left panel of fig.~\ref{fig:results} displays the posterior credible intervals obtained from this model, revealing that at N2LO the missing-higher-order uncertainty is significantly larger than that inferred from scale variation alone.
Notably, the N3LO result (purple dashed line) lies near the middle of the N2LO uncertainty band (orange shaded for 68\%, and orange dotted lines for 95\%), demonstrating that the uncertainty estimate at N2LO is consistent with the observed size of the N3LO correction.
The median of the distribution lies slightly away from the perturbative result, as the model learns the alternating-sign pattern of the series.

Conditioning the model on all known perturbative coefficients through a given perturbative order yields the corresponding posterior.
To obtain our final uncertainty estimates at N2LO and N3LO, we combine the posterior distributions obtained at different values of $X$ using the scale-marginalization prescription of Ref.~\cite{Duhr:2021mfd}, thus accounting for both missing-higher-order and renormalization-scale uncertainties.
The resulting pressure and credible intervals are finally plotted as functions of $\alpha_e(X=1)$ in the right panel of fig.~\ref{fig:results}.

\section{Discussion and outlook}

In this Letter, we have presented the first four-loop result for a thermodynamic quantity in a physical gauge theory --- the next-to-next-to-next-to-leading-order pressure of a cold and dense electron gas --- which represents the first new perturbative order for the quantity in nearly 50 years \cite{Freedman:1976dm,Freedman:1976ub}.
This advance builds on a series of earlier works addressing two complementary challenges: The effective-theory treatment of infrared-divergent contributions \cite{Kurkela:2016was,Gorda:2022zyc,Gorda:2018gpy,Gorda:2021znl,Gorda:2023mkk} and the technical evaluation of four-loop diagrams at finite density \cite{Navarrete:2024zgz,Karkkainen:2025nkz}.
The key new ingredient enabling the completion of this calculation was the application of thermal Loop Tree Duality to the four-loop integrals that had previously resisted evaluation.

While the direct phenomenological impact of the new result is limited by the small value of the fine-structure constant, its convergence and renormalization-scale dependence provide a useful benchmark for the corresponding strong-interaction calculation.
Our main finding, based on a careful analysis following the strategy of \cite{Duhr:2021mfd,Gorda:2023usm}, is that the uncertainty associated with missing higher-order terms in the weak-coupling expansion of the QED pressure is likely significantly larger than that inferred from the residual scale dependence of the result.
Despite this, the overall convergence properties of the pressure appear very good, with successive perturbative corrections remaining under control and the expansion retaining predictive power up to couplings $\alpha_e \approx 0.4$.

Finally, the present work makes an important contribution to the determination of the N3LO pressure of cold quark matter by resolving several of the integrals required for the non-Abelian case.
Completing the QCD calculation will eventually allow us to compare the convergence properties of the two theories and use the resulting pressure to constrain the equation of state of neutron-star matter over a broad range of densities \cite{Annala:2023cwx,Gorda:2025aiu}.

\emph{Acknowledgements.}---%
We thank Tyler Gorda, Keijo Kajantie, Guy Moore, Saga S\"appi, York Schr\"oder, and Juuso \"Osterman for useful discussions.
The work has been supported by the Ram\'on y Cajal program under grant RYC2024-048348-I (SA), the Research Council of Finland through projects 354533 (AV) and 347499 (RP) as well as the Centre of Excellence in Neutron-Star Physics, project 374062 (AV); the European Research Council projects ExPertQCD, grant No.~101231521 (RP) and JANUS, grant No.~101044599 (MD); the Research Council of Norway through the FRIPRO programme (CoreQCD, project number 361873; AK); and the National Science Foundation (NSF) through grant PHY-2309135 to the Kavli Institute for Theoretical Physics. 
Computational resources have been provided by CSC - IT Center for Science, Finland.

\emph{AI statement.}---%
ChatGPT (OpenAI) was used for stylistic editing and to improve the presentation and discussion of the results. No AI tools were used to perform or validate the calculations.

\nocite{Gynther:2007bw,Smirnov:2010hn}
\bibliography{refPRL.bib}

@article{Tkachov:1981wb,
    author = "Tkachov, F. V.",
    title = "{A theorem on analytical calculability of 4-loop renormalization group functions}",
    doi = "10.1016/0370-2693(81)90288-4",
    journal = "Phys. Lett. B",
    volume = "100",
    pages = "65--68",
    year = "1981"
}

@article{Kurkela:2016was,
    author = "Kurkela, Aleksi and Vuorinen, Aleksi",
    title = "{Cool quark matter}",
    eprint = "1603.00750",
    archivePrefix = "arXiv",
    primaryClass = "hep-ph",
    reportNumber = "HIP-2016-06-TH, CERN-TH-2016-045",
    doi = "10.1103/PhysRevLett.117.042501",
    journal = "Phys. Rev. Lett.",
    volume = "117",
    number = "4",
    pages = "042501",
    year = "2016"
}

@article{Cacciari:2011ze,
    author = "Cacciari, Matteo and Houdeau, Nicolas",
    title = "{Meaningful characterisation of perturbative theoretical uncertainties}",
    eprint = "1105.5152",
    archivePrefix = "arXiv",
    primaryClass = "hep-ph",
    doi = "10.1007/JHEP09(2011)039",
    journal = "JHEP",
    volume = "09",
    pages = "039",
    year = "2011"
}

@article{Duhr:2021mfd,
    author = "Duhr, Claude and Huss, Alexander and Mazeliauskas, Aleksas and Szafron, Robert",
    title = "{An analysis of Bayesian estimates for missing higher orders in perturbative calculations}",
    eprint = "2106.04585",
    archivePrefix = "arXiv",
    primaryClass = "hep-ph",
    reportNumber = "CERN-TH-2021-058",
    doi = "10.1007/JHEP09(2021)122",
    journal = "JHEP",
    volume = "09",
    pages = "122",
    year = "2021"
}

@book{Bellac:2011kqa,
    author = "Bellac, Michel Le",
    title = "{Thermal Field Theory}",
    doi = "10.1017/CBO9780511721700",
    isbn = "978-0-511-88506-8, 978-0-521-65477-7",
    publisher = "Cambridge University Press",
    series = "Cambridge Monographs on Mathematical Physics",
    month = "3",
    year = "2011"
}

@article{Annala:2023cwx,
    author = {Annala, Eemeli and Gorda, Tyler and Hirvonen, Joonas and Komoltsev, Oleg and Kurkela, Aleksi and N{\"a}ttil{\"a}, Joonas and Vuorinen, Aleksi},
    title = "{Strongly interacting matter exhibits deconfined behavior in massive neutron stars}",
    eprint = "2303.11356",
    archivePrefix = "arXiv",
    primaryClass = "astro-ph.HE",
    reportNumber = "HIP-2023-5/TH, HIP-2023-5/TH",
    doi = "10.1038/s41467-023-44051-y",
    journal = "Nature Commun.",
    volume = "14",
    number = "1",
    pages = "8451",
    year = "2023"
}

@article{Gorda:2023usm,
    author = "Gorda, Tyler and Komoltsev, Oleg and Kurkela, Aleksi and Mazeliauskas, Aleksas",
    title = "{Bayesian uncertainty quantification of perturbative QCD input to the neutron-star equation of state}",
    eprint = "2303.02175",
    archivePrefix = "arXiv",
    primaryClass = "hep-ph",
    doi = "10.1007/JHEP06(2023)002",
    journal = "JHEP",
    volume = "06",
    pages = "002",
    year = "2023"
}

@article{Ghisoiu:2016swa,
    author = {Ghisoiu, Ioan and Gorda, Tyler and Kurkela, Aleksi and Romatschke, Paul and S{\"a}ppi, Saga and Vuorinen, Aleksi},
    title = "{On high-order perturbative calculations at finite density}",
    eprint = "1609.04339",
    archivePrefix = "arXiv",
    primaryClass = "hep-ph",
    reportNumber = "HIP-2016-25-TH, CERN-TH-2016-207",
    doi = "10.1016/j.nuclphysb.2016.11.023",
    journal = "Nucl. Phys. B",
    volume = "915",
    pages = "102--118",
    year = "2017"
}

@article{Kurkela:2009gj,
    author = "Kurkela, Aleksi and Romatschke, Paul and Vuorinen, Aleksi",
    title = "{Cold Quark Matter}",
    eprint = "0912.1856",
    archivePrefix = "arXiv",
    primaryClass = "hep-ph",
    reportNumber = "BI-TP-2009-30, CERN-PH-TH-2009-229, INT-PUB-09-060, TUW-09-19",
    doi = "10.1103/PhysRevD.81.105021",
    journal = "Phys. Rev. D",
    volume = "81",
    pages = "105021",
    year = "2010"
}

@article{Smirnov:2010hn,
    author = "Smirnov, A. V. and Petukhov, A. V.",
    title = "{The Number of Master Integrals is Finite}",
    eprint = "1004.4199",
    archivePrefix = "arXiv",
    primaryClass = "hep-th",
    doi = "10.1007/s11005-010-0450-0",
    journal = "Lett. Math. Phys.",
    volume = "97",
    pages = "37--44",
    year = "2011"
}

@article{Gorda:2018gpy,
    author = {Gorda, Tyler and Kurkela, Aleksi and Romatschke, Paul and S{\"a}ppi, Saga and Vuorinen, Aleksi},
    title = "{Next-to-Next-to-Next-to-Leading Order Pressure of Cold Quark Matter: Leading Logarithm}",
    eprint = "1807.04120",
    archivePrefix = "arXiv",
    primaryClass = "hep-ph",
    reportNumber = "CERN-TH-2018-230, HIP-2018-13/TH",
    doi = "10.1103/PhysRevLett.121.202701",
    journal = "Phys. Rev. Lett.",
    volume = "121",
    number = "20",
    pages = "202701",
    year = "2018"
}

@article{Gorda:2021znl,
    author = {Gorda, Tyler and Kurkela, Aleksi and Paatelainen, Risto and S{\"a}ppi, Saga and Vuorinen, Aleksi},
    title = "{Soft Interactions in Cold Quark Matter}",
    eprint = "2103.05658",
    archivePrefix = "arXiv",
    primaryClass = "hep-ph",
    reportNumber = "HIP-2021-9/TH",
    doi = "10.1103/PhysRevLett.127.162003",
    journal = "Phys. Rev. Lett.",
    volume = "127",
    number = "16",
    pages = "162003",
    year = "2021"
}

@article{Gorda:2021kme,
    author = {Gorda, Tyler and Kurkela, Aleksi and Paatelainen, Risto and S{\"a}ppi, Saga and Vuorinen, Aleksi},
    title = "{Cold quark matter at N3LO: Soft contributions}",
    eprint = "2103.07427",
    archivePrefix = "arXiv",
    primaryClass = "hep-ph",
    reportNumber = "HIP-2021-10/TH",
    doi = "10.1103/PhysRevD.104.074015",
    journal = "Phys. Rev. D",
    volume = "104",
    number = "7",
    pages = "074015",
    year = "2021"
}

@article{Navarrete:2024ruu,
    author = {Navarrete, Pablo and Schr{\"o}der, York},
    title = "{The g$^{6}$ pressure of hot Yang-Mills theory: canonical form of the integrand}",
    eprint = "2408.15830",
    archivePrefix = "arXiv",
    primaryClass = "hep-ph",
    doi = "10.1007/JHEP11(2024)037",
    journal = "JHEP",
    volume = "11",
    pages = "037",
    year = "2024"
}

@article{Moore:2023glb,
    author = "Moore, Guy D. and Gorda, Tyler",
    title = "{Bounding the QCD Equation of State with the Lattice}",
    eprint = "2309.15149",
    archivePrefix = "arXiv",
    primaryClass = "nucl-th",
    doi = "10.1007/JHEP12(2023)133",
    journal = "JHEP",
    volume = "12",
    pages = "133",
    year = "2023"
}

@article{Vuorinen:2003fs,
    author = "Vuorinen, A.",
    title = "{The Pressure of QCD at finite temperatures and chemical potentials}",
    eprint = "hep-ph/0305183",
    archivePrefix = "arXiv",
    reportNumber = "HIP-2003-30-TH",
    doi = "10.1103/PhysRevD.68.054017",
    journal = "Phys. Rev. D",
    volume = "68",
    pages = "054017",
    year = "2003"
}

@article{Freedman:1976dm,
    author = "Freedman, Barry A. and McLerran, Larry D.",
    title = "{Fermions and Gauge Vector Mesons at Finite Temperature and Density. 2. The Ground State Energy of a Relativistic electron Gas}",
    reportNumber = "MIT-CTP-576",
    doi = "10.1103/PhysRevD.16.1147",
    journal = "Phys. Rev. D",
    volume = "16",
    pages = "1147",
    year = "1977"
}

@article{Freedman:1976ub,
    author = "Freedman, Barry A. and McLerran, Larry D.",
    title = "{Fermions and Gauge Vector Mesons at Finite Temperature and Density. 3. The Ground State Energy of a Relativistic Quark Gas}",
    reportNumber = "MIT-CTP-582",
    doi = "10.1103/PhysRevD.16.1169",
    journal = "Phys. Rev. D",
    volume = "16",
    pages = "1169",
    year = "1977"
}

@article{Navarrete:2024zgz,
    author = {Navarrete, Pablo and Paatelainen, Risto and Sepp{\"a}nen, Kaapo},
    title = "{Perturbative QCD meets phase quenching: The pressure of cold quark matter}",
    eprint = "2403.02180",
    archivePrefix = "arXiv",
    primaryClass = "hep-ph",
    reportNumber = "HIP-2024-5/TH",
    doi = "10.1103/PhysRevD.110.094033",
    journal = "Phys. Rev. D",
    volume = "110",
    number = "9",
    pages = "094033",
    year = "2024"
}

@article{Gorda:2022zyc,
    author = {Gorda, Tyler and Kurkela, Aleksi and {\"O}sterman, Juuso and Paatelainen, Risto and S{\"a}ppi, Saga and Schicho, Philipp and Sepp{\"a}nen, Kaapo and Vuorinen, Aleksi},
    title = "{Degenerate fermionic matter at N3LO: Quantum electrodynamics}",
    eprint = "2204.11893",
    archivePrefix = "arXiv",
    primaryClass = "hep-ph",
    reportNumber = "HIP-2022-9/TH",
    doi = "10.1103/PhysRevD.107.L031501",
    journal = "Phys. Rev. D",
    volume = "107",
    number = "3",
    pages = "L031501",
    year = "2023"
}

@article{Gorda:2025cwu,
    author = {Gorda, Tyler and Navarrete, Pablo and Paatelainen, Risto and Sandbote, Leon and Sepp{\"a}nen, Kaapo},
    title = "{New approach to determine the thermodynamics of deconfined matter to high accuracy}",
    eprint = "2511.09627",
    archivePrefix = "arXiv",
    primaryClass = "hep-ph",
    reportNumber = "HIP-2025-32/TH",
    doi = "10.1103/wd4z-cqw3",
    journal = "Phys. Rev. D",
    volume = "114",
    number = "5",
    pages = "054056",
    year = "2026"
}

@unpublished{Davies:2026cci,
    author = "Davies, J. and Kaneko, T. and Marinissen, C. and Ueda, T. and Vermaseren, J. A. M.",
    title = "{FORM Version 5.0}",
    eprint = "2601.19982",
    archivePrefix = "arXiv",
    primaryClass = "hep-ph",
    month = "1",
    year = "2026"
}

@article{Nogueira:1991ex,
    author = "Nogueira, Paulo",
    title = "{Automatic Feynman Graph Generation}",
    reportNumber = "IFM-7-91",
    doi = "10.1006/jcph.1993.1074",
    journal = "J. Comput. Phys.",
    volume = "105",
    pages = "279--289",
    year = "1993"
}

@article{Osterman:2023tnt,
    author = {{\"O}sterman, Juuso and Schicho, Philipp and Vuorinen, Aleksi},
    title = "{Integrating by parts at finite density}",
    eprint = "2304.05427",
    archivePrefix = "arXiv",
    primaryClass = "hep-ph",
    reportNumber = "HIP-2023-7/TH",
    doi = "10.1007/JHEP08(2023)212",
    journal = "JHEP",
    volume = "08",
    pages = "212",
    year = "2023"
}

@article{Nishimura:2012ee,
    author = "Nishimura, M. and Schroder, Y.",
    title = "{IBP methods at finite temperature}",
    eprint = "1207.4042",
    archivePrefix = "arXiv",
    primaryClass = "hep-ph",
    reportNumber = "BI-TP-2012-28",
    doi = "10.1007/JHEP09(2012)051",
    journal = "JHEP",
    volume = "09",
    pages = "051",
    year = "2012"
}

@article{Gorda:2023mkk,
    author = {Gorda, Tyler and Paatelainen, Risto and S{\"a}ppi, Saga and Sepp{\"a}nen, Kaapo},
    title = "{Equation of State of Cold Quark Matter to $O(\alpha_s^3 \ln \alpha_s)$}",
    eprint = "2307.08734",
    archivePrefix = "arXiv",
    primaryClass = "hep-ph",
    reportNumber = "HIP-2023-10/TH, TUM-EFT 181/23",
    doi = "10.1103/PhysRevLett.131.181902",
    journal = "Phys. Rev. Lett.",
    volume = "131",
    number = "18",
    pages = "181902",
    year = "2023"
}

@article{Karkkainen:2025nkz,
    author = {K{\"a}rkk{\"a}inen, Aapeli and Navarrete, Pablo and Nurmela, Mika and Paatelainen, Risto and Sepp{\"a}nen, Kaapo and Vuorinen, Aleksi},
    title = "{Quark Matter at Four Loops: Hardships and How to Overcome Them}",
    eprint = "2501.17921",
    archivePrefix = "arXiv",
    primaryClass = "hep-ph",
    reportNumber = "HIP-2025-01/TH",
    doi = "10.1103/627n-5g6l",
    journal = "Phys. Rev. Lett.",
    volume = "135",
    number = "2",
    pages = "021901",
    year = "2025"
}

@article{Chetyrkin:1981qh,
    author = "Chetyrkin, K. G. and Tkachov, F. V.",
    title = "{Integration by parts: The algorithm to calculate $\beta$-functions in 4 loops}",
    doi = "10.1016/0550-3213(81)90199-1",
    journal = "Nucl. Phys. B",
    volume = "192",
    pages = "159--204",
    year = "1981"
}

@phdthesis{Seppanen:2025owq,
    author = {Sepp{\"a}nen, Kaapo},
    title = "{Quark Matter Thermodynamics from High-Order Perturbative QCD}",
    school = "Helsinki U.",
    year = "2025"
}

@article{Remiddi:1997ny,
    author = "Remiddi, Ettore",
    title = "{Differential equations for Feynman graph amplitudes}",
    eprint = "hep-th/9711188",
    archivePrefix = "arXiv",
    reportNumber = "DFUB-97-15, DFUB 97-15",
    doi = "10.1007/BF03185566",
    journal = "Nuovo Cim. A",
    volume = "110",
    pages = "1435--1452",
    year = "1997"
}

@article{Gehrmann:1999as,
    author = "Gehrmann, T. and Remiddi, E.",
    title = "{Differential equations for two-loop four-point functions}",
    eprint = "hep-ph/9912329",
    archivePrefix = "arXiv",
    reportNumber = "TTP-99-49",
    doi = "10.1016/S0550-3213(00)00223-6",
    journal = "Nucl. Phys. B",
    volume = "580",
    pages = "485--518",
    year = "2000"
}

@article{Capatti:2019ypt,
    author = "Capatti, Zeno and Hirschi, Valentin and Kermanschah, Dario and Ruijl, Ben",
    title = "{Loop-Tree Duality for Multiloop Numerical Integration}",
    eprint = "1906.06138",
    archivePrefix = "arXiv",
    primaryClass = "hep-ph",
    doi = "10.1103/PhysRevLett.123.151602",
    journal = "Phys. Rev. Lett.",
    volume = "123",
    number = "15",
    pages = "151602",
    year = "2019"
}

@article{Gorda:2022fci,
    author = {Gorda, Tyler and Kurkela, Aleksi and {\"O}sterman, Juuso and Paatelainen, Risto and S{\"a}ppi, Saga and Schicho, Philipp and Sepp{\"a}nen, Kaapo and Vuorinen, Aleksi},
    title = "{Soft photon propagation in a hot and dense medium to next-to-leading order}",
    eprint = "2204.11279",
    archivePrefix = "arXiv",
    primaryClass = "hep-ph",
    reportNumber = "HIP-2022-8/TH",
    doi = "10.1103/PhysRevD.107.036012",
    journal = "Phys. Rev. D",
    volume = "107",
    number = "3",
    pages = "036012",
    year = "2023"
}

@article{Laporta:2000dsw,
    author = "Laporta, S.",
    title = "{High-precision calculation of multiloop Feynman integrals by difference equations}",
    eprint = "hep-ph/0102033",
    archivePrefix = "arXiv",
    doi = "10.1142/S0217751X00002159",
    journal = "Int. J. Mod. Phys. A",
    volume = "15",
    pages = "5087--5159",
    year = "2000"
}

@book{GiulianiVignale2005,
  author    = {G. F. Giuliani and G. Vignale},
  title     = {Quantum Theory of the Electron Liquid},
  publisher = {Cambridge University Press},
  address   = {Cambridge},
  year      = {2005},
  isbn      = {978-0521821124}
}

@article{Gynther:2007bw,
    author = "Gynther, A. and Laine, M. and Schroder, Y. and Torrero, C. and Vuorinen, A.",
    title = "{Four-loop pressure of massless O(N) scalar field theory}",
    eprint = "hep-ph/0703307",
    archivePrefix = "arXiv",
    reportNumber = "BI-TP-2007-05, ECT-07-06",
    doi = "10.1088/1126-6708/2007/04/094",
    journal = "JHEP",
    volume = "04",
    pages = "094",
    year = "2007"
}

@article{Gorda:2025aiu,
    author = "Gorda, Tyler and Komoltsev, Oleg and Kurkela, Aleksi and Sunde, Eirik",
    title = "{Constrained Gaussian-process-bridge Prior for Neutron-star Equation-of-state Inference}",
    eprint = "2512.18044",
    archivePrefix = "arXiv",
    primaryClass = "astro-ph.HE",
    doi = "10.3847/1538-4357/ae552a",
    journal = "Astrophys. J.",
    volume = "1002",
    number = "1",
    pages = "40",
    year = "2026"
}

@unpublished{companion,
    author = "{A.~K{\"a}rkk{\"a}inen, P.~Navarrete, M.~Nurmela, R.~Paatelainen, Y.~Schr{\"o}der, and K.~Sepp{\"a}nen}",
    note = "in preparation (2026)"
}

@misc{Ruijl:Symbolica:2026,
    author = {Ruijl, Ben},
    title = {{Symbolica (3.0.1)}},
    year = {2026},
    publisher = {Ruijl Research},
    doi = {10.5281/zenodo.23180832},
    url = {https://doi.org/10.5281/zenodo.23180832},
    note = {\url{https://doi.org/10.5281/zenodo.23180832}}
}

@article{Navarrete:2025yxy,
    author = {Navarrete, Pablo and Paatelainen, Risto and Sepp{\"a}nen, Kaapo and Tenkanen, Tuomas V. I.},
    title = "{Cosmological phase transitions without high-temperature expansions}",
    eprint = "2507.07014",
    archivePrefix = "arXiv",
    primaryClass = "hep-ph",
    reportNumber = "HIP-2025-20/TH",
    doi = "10.1007/JHEP01(2026)113",
    journal = "JHEP",
    volume = "2026",
    number = "01",
    pages = "113",
    year = "2026"
}

@article{Raza:2026imv,
    author = "Raza, Tousif and Gorda, Tyler",
    title = "{Stability and Structural Properties of Hot Quark Stars within Perturbative QCD}",
    eprint = "2609.14768",
    archivePrefix = "arXiv",
    primaryClass = "nucl-th",
    journal = {},
    month = "9",
    year = "2026"
}

@article{Catani:2008xa,
    author = "Catani, Stefano and Gleisberg, Tanju and Krauss, Frank and Rodrigo, German and Winter, Jan-Christopher",
    title = "{From Loops to Trees by-Passing Feynman's Theorem}",
    eprint = "0804.3170",
    archivePrefix = "arXiv",
    primaryClass = "hep-ph",
    reportNumber = "FERMILAB-PUB-08-092-T, IFIC-08-21, IPPP-08-22, SLAC-PUB-13218",
    doi = "10.1088/1126-6708/2008/09/065",
    journal = "JHEP",
    volume = "09",
    pages = "065",
    year = "2008"
}

@article{Bierenbaum:2010cy,
    author = "Bierenbaum, Isabella and Catani, Stefano and Draggiotis, Petros and Rodrigo, German",
    title = "{A Tree-Loop Duality Relation at Two Loops and Beyond}",
    eprint = "1007.0194",
    archivePrefix = "arXiv",
    primaryClass = "hep-ph",
    reportNumber = "IFIC-10-17",
    doi = "10.1007/JHEP10(2010)073",
    journal = "JHEP",
    volume = "10",
    pages = "073",
    year = "2010"
}

@article{Runkel:2019yrs,
    author = "Runkel, Robert and Sz{\H{o}}r, Zolt{\'a}n and Vesga, Juan Pablo and Weinzierl, Stefan",
    title = "{Causality and loop-tree duality at higher loops}",
    eprint = "1902.02135",
    archivePrefix = "arXiv",
    primaryClass = "hep-ph",
    doi = "10.1103/PhysRevLett.122.111603",
    journal = "Phys. Rev. Lett.",
    volume = "122",
    number = "11",
    pages = "111603",
    year = "2019",
    note = "[Erratum: Phys.Rev.Lett. 123, 059902 (2019)]"
}

@article{Capatti:2020ytd,
    author = "Capatti, Zeno and Hirschi, Valentin and Kermanschah, Dario and Pelloni, Andrea and Ruijl, Ben",
    title = "{Manifestly Causal Loop-Tree Duality}",
    eprint = "2009.05509",
    archivePrefix = "arXiv",
    primaryClass = "hep-ph",
    journal = {},
    month = "9",
    year = "2020"
}

@article{Capatti:2022mly,
    author = "Capatti, Zeno",
    title = "{Exposing the threshold structure of loop integrals}",
    eprint = "2211.09653",
    archivePrefix = "arXiv",
    primaryClass = "hep-th",
    doi = "10.1103/PhysRevD.107.L051902",
    journal = "Phys. Rev. D",
    volume = "107",
    number = "5",
    pages = "L051902",
    year = "2023"
}

@article{JesusAguilera-Verdugo:2020fsn,
    author = "Jes{\'u}s Aguilera-Verdugo, J. and Hern{\'a}ndez-Pinto, Roger J. and Rodrigo, Germ{\'a}n and Sborlini, German F. R. and Torres Bobadilla, William J.",
    title = "{Mathematical properties of nested residues and their application to multi-loop scattering amplitudes}",
    eprint = "2010.12971",
    archivePrefix = "arXiv",
    primaryClass = "hep-ph",
    reportNumber = "IFIC/20-30; DESY 20-172; MPP-2020-184",
    doi = "10.1007/JHEP02(2021)112",
    journal = "JHEP",
    volume = "02",
    pages = "112",
    year = "2021"
}

@article{Aguilera-Verdugo:2020set,
    author = "Aguilera-Verdugo, J. Jesus and Driencourt-Mangin, Felix and Hern{\'a}ndez-Pinto, Roger J. and Plenter, Judith and Ramirez-Uribe, Selomit and Renteria Olivo, Andres E. and Rodrigo, German and Sborlini, German F. R. and Torres Bobadilla, William J. and Tracz, Szymon",
    title = "{Open Loop Amplitudes and Causality to All Orders and Powers from the Loop-Tree Duality}",
    eprint = "2001.03564",
    archivePrefix = "arXiv",
    primaryClass = "hep-ph",
    reportNumber = "IFIC/20-02",
    doi = "10.1103/PhysRevLett.124.211602",
    journal = "Phys. Rev. Lett.",
    volume = "124",
    number = "21",
    pages = "211602",
    year = "2020"
}

@article{Herzog:2017ohr,
    author = "Herzog, F. and Ruijl, B. and Ueda, T. and Vermaseren, J. A. M. and Vogt, A.",
    title = "{The five-loop beta function of Yang-Mills theory with fermions}",
    eprint = "1701.01404",
    archivePrefix = "arXiv",
    primaryClass = "hep-ph",
    reportNumber = "NIKHEF-2017-001, LTH-1117",
    doi = "10.1007/JHEP02(2017)090",
    journal = "JHEP",
    volume = "02",
    pages = "090",
    year = "2017"
}

@article{Kotikov:1990kg,
    author = "Kotikov, A. V.",
    title = "{Differential equations method: New technique for massive Feynman diagrams calculation}",
    reportNumber = "ITF-90-31E",
    doi = "10.1016/0370-2693(91)90413-K",
    journal = "Phys. Lett. B",
    volume = "254",
    pages = "158--164",
    year = "1991"
}

@unpublished{newapp,
    author = "Abreu, Samuel and Delto, Maximilian and Kurkela, Aleksi",
    note = "in preparation (2026)"
}

@article{Tarasov:1997kx,
    author = "Tarasov, O. V.",
    title = "{Generalized recurrence relations for two loop propagator integrals with arbitrary masses}",
    eprint = "hep-ph/9703319",
    archivePrefix = "arXiv",
    reportNumber = "DESY-97-065, JINR-E2-96-433",
    doi = "10.1016/S0550-3213(97)00376-3",
    journal = "Nucl. Phys. B",
    volume = "502",
    pages = "455--482",
    year = "1997"
}

@article{Arnold:1994ps,
    author = "Arnold, Peter Brockway and Zhai, Cheng-Xing",
    title = "{The Three loop free energy for pure gauge QCD}",
    eprint = "hep-ph/9408276",
    archivePrefix = "arXiv",
    reportNumber = "UW-PT-94-03",
    doi = "10.1103/PhysRevD.50.7603",
    journal = "Phys. Rev. D",
    volume = "50",
    pages = "7603--7623",
    year = "1994"
}

@article{Arnold:1994eb,
    author = "Arnold, Peter Brockway and Zhai, Cheng-xing",
    title = "{The Three loop free energy for high temperature QED and QCD with fermions}",
    eprint = "hep-ph/9410360",
    archivePrefix = "arXiv",
    reportNumber = "UW-PT-94-11",
    doi = "10.1103/PhysRevD.51.1906",
    journal = "Phys. Rev. D",
    volume = "51",
    pages = "1906--1918",
    year = "1995"
}

@article{Baikov:2016tgj,
    author        = {Baikov, P. A. and Chetyrkin, K. G. and K{\"u}hn, J. H.},
    title         = {{Five-Loop Running of the QCD Coupling Constant}},
    journal       = {Phys. Rev. Lett.},
    volume        = {118},
    pages         = {082002},
    year          = {2017},
    doi           = {10.1103/PhysRevLett.118.082002},
    eprint        = {1606.08659},
    archivePrefix = {arXiv},
    primaryClass  = {hep-ph}
}

@article{Luthe:2016xec,
    author        = {Luthe, Thomas and Maier, Andreas and
                     Marquard, Peter and Schr{\"o}der, York},
    title         = {{Five-loop quark mass and field anomalous
                      dimensions for a general gauge group}},
    journal       = {JHEP},
    volume        = {01},
    pages         = {081},
    year          = {2017},
    doi           = {10.1007/JHEP01(2017)081},
    eprint        = {1612.05512},
    archivePrefix = {arXiv},
    primaryClass  = {hep-ph}
}

@article{Luthe:2017ttc,
    author        = {Luthe, Thomas and Maier, Andreas and
                     Marquard, Peter and Schr{\"o}der, York},
    title         = {{Complete renormalization of QCD at five loops}},
    journal       = {JHEP},
    volume        = {03},
    pages         = {020},
    year          = {2017},
    doi           = {10.1007/JHEP03(2017)020},
    eprint        = {1701.07068},
    archivePrefix = {arXiv},
    primaryClass  = {hep-ph}
}

\section*{End matter}

\appendix

\section{Master integrals}\label{app:masters}

In this appendix, we elaborate on the integral structures that contribute to the four-loop diagrams entering the loop expansion of the QED pressure. A more detailed discussion will be presented in a forthcoming paper \cite{companion}.

After all graphs in fig.~\ref{fig:diags} are scalarized and reduced into a canonical form, we obtain a set of altogether 75 four-loop master integrals. 
The list representation we employ uses ten quadratic denominators $P_n^2\equiv P_n\cdot P_n$, constructed from four independent loop momenta $P_1,\dots,P_4$. 
Following \cite{Karkkainen:2025nkz}, the remaining linear combinations are defined by
\begin{equation}
    (P_5,\dots,P_{10})\equiv(P_{14},P_{24},P_{34},P_{12},P_{13},P_{123}),
\end{equation}
where $P_{n m}\equiv P_{n}-P_{m}$ and $P_{123}\equiv P_{12}-P_3$. 
A four-loop vacuum integral contributing to $p^\text{h}$ is then specified by a list of ten integers $(s_1,\dots,s_{10})$, with a positive entry $s_j$ denoting a denominator factor $(P_j^2)^{s_j}$. 
Subscripts on the first four entries specify the shifts $p_{j,0}\to p_{j,0}+i\mu\eta_j$, with $+$, $-$, and $++$ denoting $\eta_j=1,-1$, and $2$, respectively, while an omitted subscript denotes zero shift. 
These labels apply to the loop momenta wherever they occur, even when the corresponding entry is zero. 

Unlike in \cite{Karkkainen:2025nkz}, we represent numerators using scalar products, with a negative entry $s_j$, $j=5,\dots,10$, denoting the power $-s_j$ of the corresponding product in the ordered set
\begin{equation}
    (P_1\cdot P_4,P_2\cdot P_4,P_3\cdot P_4,P_1\cdot P_2,P_1\cdot P_3,P_2\cdot P_3).
\end{equation}
As one example of this notation, one master integral appearing in the pressure of QED at four loops reads
\begin{equation}
\label{eq:listex}
    (1,1_+,1_-,1,1,1,1,1,1,-3)=\int_{P_1P_2^+P_3^-P_4} \frac{(P_2 \cdot P_3)^3}{P_1^2 \dots P_9^2},
\end{equation}
where we have employed the bosonic integration measure
\begin{equation}
    \int_{P} \equiv \left( \frac{e^{\gamma_\text{E}}\bar{\Lambda}^2}{4\pi} \right)^{\frac{3-d}{2}}\int_{-\infty}^{\infty}\frac{\text{d}p_0}{2\pi} \int \frac{\text{d}^d\mathbf{p}}{(2\pi)^d},
\end{equation}
while in the fermionic case we simply shift the temporal component in accordance with the superscript in $P$.

\begin{figure}[t]
    \centering
    \includegraphics[width=0.95\linewidth]{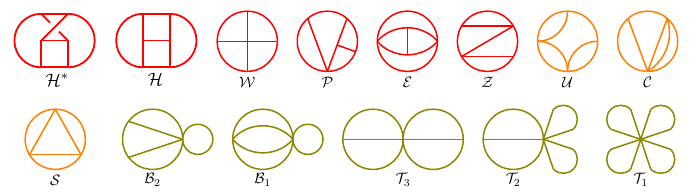}
    \caption{Four-loop vacuum topologies contributing to the pressure of QED as defined e.g. in \cite{Karkkainen:2025nkz} (see also \cite{Navarrete:2024ruu}). 
    The color coding reflects the complexity of the integral, with red standing for the most challenging and olive for factorized structures.}
    \label{fig:topos}
\end{figure}

As explained in \cite{Karkkainen:2025nkz} (see \cite{Navarrete:2024ruu} for a more precise definition), a master integral belongs to a certain class of integrals defined by its topology, which provides useful information on their sub-structure. The vacuum topologies required for QED are shown in fig.~\ref{fig:topos}, with the colors chosen to roughly convey the complexity of the associated integrals given the toolkit available.

Starting from the simplest entities, the graphs in olive factorize into lower-loop structures and can therefore be tackled with semi-analytic techniques that have proven successful at three loops --- most notably the cutting rules of \cite{Ghisoiu:2016swa}. 
We have cross-checked several cases in this class of integrals against other methods, including IBP reductions \cite{Osterman:2023tnt} and self-energy manipulations \cite{Vuorinen:2003fs}.

The non-factorized topologies shown in orange also have comparatively simple structures, featuring one-loop self-energy insertions or fewer propagators, but nevertheless lie near the practical limits of our current semi-analytic methods. 
Two of the simpler cases in this class were already evaluated in \cite{Karkkainen:2025nkz} and cross-checked against numerical results obtained with LTD. 

Finally, the red topologies contain the most challenging integrals, particularly $\mathcal{H}$, $\mathcal{W}$, and the non-planar $\mathcal{H}^*$, which is absent in Yang--Mills theory \cite{Navarrete:2024ruu}. For these integrals, we rely on expensive numerical calculations employing the LTD approach \cite{companion}.

As explained in some detail in  \cite{Seppanen:2025owq}, the LTD approach amounts to integrating over the loop energies analytically, with subtractions of local UV counterterms rendering the integrals suitable for numerical treatment.
We evaluate the remaining spatial integrals by Monte-Carlo integration using our LTD setup, which relies on the use of \texttt{Symbolica} \cite{Ruijl:Symbolica:2026}.
The UV counterterms are evaluated analytically in dimensional regularization and added back, yielding the $\varepsilon$ expansion of each master integral. 
As a technical sidenote, the $\mathcal{H}$ masters feature IR divergences, which we treat with local IR counterterms; however, these divergences are seen to cancel in the QED pressure, as illustrated in appendix~\ref{app:example}. 
Consequently, the coefficient of the IR pole in eq.~\eqref{eq:phardcoeffs} receives contributions only from topologies of complexity at most that of $\mathcal{U}$, and can therefore be extracted by analytic means.

Finally, we note that we treat all master integrals individually. 
For those computed with more than one method, we compare the expansions, including both the poles and finite terms. 
All comparisons involving numerical results agree within two standard deviations, while IBP reduction provides additional fully analytic consistency checks for a limited subset of non-trivial cases.

\begin{table*}[t]
\centering
\begin{tabular}{||c|c|c|c|c||c|}
    \hline
    $n$ & List representation & $\varepsilon^{-2}$ & $\varepsilon^{-1}$ & $\varepsilon^0$ & $c_n(d)$ \\
    \hline \hline
    1 & $(1_+,1_+,1_+,1_+,0,1,0,0,1,0)_{\mathcal{T}_3}$ & $16$ & $192$ & $-16 (-84 + \pi^2)$ & $-\frac{1}{3}(-1+15d)$ \\
    2 & $(1_+,1_+,1,0,1,1,1,0,0,0)_{\mathcal{S}}$ & $16$ & $-\frac{64}{3}(-11+2\ln 2)$ & $1407.75(5)$ & $-4(-9+d)$ \\
    3 & $(1_+,1_+,1_+,0,1,1,1,0,0,0)_{\mathcal{S}}$ & $36$ & $-4(-134+32\ln 2)$ & $2890.61(6)$ & $\frac{2}{3}(-5+3d)$ \\
    4 & $(1_+,1_+,1_+,0_{++},1,1,1,0,0,0)_{\mathcal{S}}$ & $36$ & $-4(-134+32\ln 2)$ & $2901.23(6)$ & $-\frac{2}{3}(-5+3d)$ \\
    5 & $(1,1_+,1_-,1_+,1,0,0,0,0,1)_{\mathcal{C}}$ & $10$ & $133$ & $912.22(5)$ & $-51+5d$ \\
    6 & $(1_+,1_+,1_+,1_+,1,0,0,0,0,1)_{\mathcal{C}}$ & $24$ & $\frac{4}{3}(245-32\ln 2)$ & $1903.24(6)$ & $\frac{1}{3}(17+9d)$ \\
    7 & $(1_+,1_+,1,1,1,1,1,-1,0,0)_{\mathcal{U}}$ & $4$ & $-\frac{8}{3}(-23+8\ln 2)$ & $309.013(34)$ & $16 (-9+d)$ \\
    8 & $(1_+,1_-,1_+,1,1,1,0,0,-1,1)_{\mathcal{E}}$ & $\frac{9}{2}$ & $\frac{227}{4}$ & $362.060(27)$ & $4 (1+d)$ \\
    9 & $(1_+,1_+,1_+,1,1,1,0,-1,0,1)_{\mathcal{E}}$ & $15$ & $\frac{1}{2}(437-128\ln 2)$ & $1093.570(29)$ & $2 (1+d)$ \\
    10 & $(1_+,1_+,1_+,1_+,1,1,-1,0,0,1)_{\mathcal{E}}$ & $0$ & $0$ & $-8.996(25)$ & $\frac{4}{3}(35+3d)$ \\
    11 & $(1_+,1_+,1_+,1_+,1,1,0,-1,0,1)_{\mathcal{E}}$ & $1$ & $20$ & $190.422(24)$ & $\frac{128}{3}$ \\
    12 & $(1_+,1_+,1_+,1_+,1,1,0,0,-1,1)_{\mathcal{E}}$ & $\frac{9}{2}$ & $\frac{113}{2}$ & $350.953(17)$ & $\frac{256}{3}$ \\
    13 & $(1,1_+,1_+,1,1,1,0,0,1,-1)_{\mathcal{Z}}$ & $\frac{1}{2}$ & $\frac{37}{4}$ & $68.911(31)$ & $64$ \\
    14 & $(1_+,1_+,1_+,1,1,1,-1,0,1,0)_{\mathcal{Z}}$ & $3$ & $\frac{1}{6}(289-64\ln 2)$ & $289.253(18)$ & $\frac{128}{3}$ \\
    15 & $(1_+,1_+,1_+,1,1,1,0,-1,1,0)_{\mathcal{Z}}$ & $\frac{9}{2}$ & $\frac{1}{12}(835-256\ln 2)$ & $367.196(15)$ & $-\frac{4}{3}(23+15d)$ \\
    16 & $(1_+,1_+,1_+,1,1,1,0,0,1,-1)_{\mathcal{Z}}$ & $\frac{3}{2}$ & $\frac{1}{12}(305-128\ln 2)$ & $121.484(17)$ & $-\frac{256}{3}$ \\
    17 & $(1_+,1_-,1_+,1,1,1,-1,-1,1,1)_{\mathcal{W}}$ & $\frac{3}{4}$ & $\frac{95}{8}$ & $98.04(4)$ & $-4(-5+3d)$ \\
    18 & $(1_+,1_-,1_+,1,1,1,0,-2,1,1)_{\mathcal{W}}$ & $13$ & $\frac{1}{3}(557-128\ln 2)$ & $1019.13(4)$ & $32$ \\
    19 & $(1_+,1_+,1_+,1,1,1,-1,-1,1,1)_{\mathcal{W}}$ & $\frac{3}{4}$ & $\frac{95}{8}$ & $95.378(35)$ & $4(-5+3d)$ \\
    20 & $(1_+,1_+,1_+,1,1,1,0,-2,1,1)_{\mathcal{W}}$ & $13$ & $\frac{1}{3}(557-128\ln 2)$ & $997.17(4)$ & $-32$ \\
    21 & $(1,1_+,1_-,1,1,1,1,1,1,-3)_{\mathcal{H}}$ & $-5$ & $\frac{1}{3}(-193+64\ln 2)-\kappa_{\rm IR}$ & $-306.84(23)$ & $-\frac{128}{3}$ \\
    22 & $(1,1_+,1_+,1,1,1,1,1,1,-3)_{\mathcal{H}}$ & $-5$ & $-51-32\ln2-\kappa_{\rm IR}$ & $-493.13(22)$ & $\frac{128}{3}$ \\
    \hline
\end{tabular}
\caption{Coefficients of various powers of $\varepsilon$ in our final results for the master integrals $I_n$ contributing to eq.~\eqref{eq:bugblatters}, listed together with the polynomial prefactors $c_n(d)$. 
The integrals are defined by their corresponding list representations as in eq.~\eqref{eq:listex}, and we have omitted an overall prefactor of $\mu^4 X^{8\varepsilon}/(4\pi)^8$ common to all four-loop integrals. 
The subscripts outside the parentheses identify the topologies in fig.~\ref{fig:topos}, while those of individual entries specify the chemical-potential shifts as explained in appendix~\ref{app:masters}. 
The common IR-pole contribution on rows 21 and 22, denoted by $\kappa_{\rm IR}=1.0655(20)$, cancels exactly in eq.~\eqref{eq:bugblatters} because $c_{21}=-c_{22}$.}
\label{tab:ints}
\end{table*}

\section{A challenging example}\label{app:example}

The sum of the two ``Bugblatter'' diagrams --- the last two $O(N_f^2)$ graphs of fig.~\ref{fig:diags} --- provides an independent check of our master-integral calculation. 
These diagrams involve a substantial subset of the four-loop integral topologies listed in fig.~\ref{fig:topos} (see table~\ref{tab:ints}) and are the most challenging to evaluate, encompassing all computational difficulties encountered in the full pressure calculation. 
After factoring out $e^6N_f^2\mu^4$, their sum, however, reduces to a finite, gauge-invariant dimensionless number in QED \cite{Moore:2023glb,Navarrete:2024zgz}.

In terms of the canonical master integrals, the sum of the Bugblatters can be expressed as
\begin{equation}
\label{eq:bugblatters}
    \frac{1}{6}\raisebox{-4mm}{\includegraphics[width=0.2\linewidth]{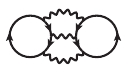}} + \frac{1}{6}\raisebox{-4mm}{\includegraphics[width=0.2\linewidth]{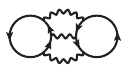}} = e^6 N_f^2 \sum_{n=1}^{22} c_n I_n,
\end{equation}
where $c_n=c_n(d)$ are linear functions of $d$ and $I_n=I_n(d,\mu)$ the four-loop master integrals, both listed in table~\ref{tab:ints}. 

Expanding the prefactors with $d=3-2\varepsilon$ and inserting the master integrals into eq.~\eqref{eq:bugblatters} gives
\begin{equation}
    \frac{1}{6}\raisebox{-4mm}{\includegraphics[width=0.2\linewidth]{figs/B1.pdf}} + \frac{1}{6}\raisebox{-4mm}{\includegraphics[width=0.2\linewidth]{figs/B2.pdf}} = e^6 N_f^2 \mu^4 \cdot b+O(\varepsilon),
\end{equation}
which demonstrates the cancellation of all poles in $\varepsilon$, including the IR poles of the two $\mathcal{H}$ masters. For the finite term, we obtain $b=-5.44(24)\cdot 10^{-7}$, which agrees within numerical uncertainties with the result of the direct diagrammatic evaluation of \cite{Navarrete:2024zgz}, $b=-5.545(12)\cdot 10^{-7}$. 
We use the more precise literature value in our final result for the pressure and consider the independent master-integral evaluation presented here a non-trivial cross-check of the full computation.

\end{document}